# *On the Upper Ocean Turbulent Dissipation Rate due to Very Short Breaking Wind-Waves*

Michael L. Banner and Russel P. Morison

School of Mathematics and Statistics,
The University of New South Wales, Sydney, Australia 2052

Corresponding author:   Michael Banner (m.banner@unsw.edu.au)

*Abstract*

Sutherland and Melville (2015a) investigated the relative contributions to the total dissipation rate in the ocean surface wave boundary layer of different breaking wave scales, from large-scale whitecaps to micro-breakers. Based on their measurements of geometric/kinematic properties of breaking waves for a wide range of wave ages, they inferred the dissipation rates from breaking as a function of scale. These results were compared with their complementary measurements of the total dissipation rate in the underlying wave boundary layer. They reported that the total depth-integrated dissipation rate in the water column agreed well with dissipation rate from breaking waves for young to very old wind seas. They also reported high observed levels of dissipation rate very near the sea surface. They concluded that this showed a large fraction of the total dissipation rate was due to non-air entraining micro-breakers and very small whitecaps. Because of its fundamental importance, both physically and for accurate air-sea interaction modeling, we assess the validity of these conclusions. Our analysis of their data shows that the microbreakers and small whitecaps do not make a significant contribution to the total turbulent dissipation rate in the wave boundary layer.

1. *Introduction*

It is well-recognised that turbulent kinetic energy (TKE) is injected sporadically at the wind-driven sea surface under active wave breaking conditions (e.g. Craig and Banner (1994), Melville (1994), Terray et al. (1996)). The breaker scale bandwidth and spatio-temporal frequencies depend on the sea state variables, primarily wind speed and wave age. Young wind seas experience a higher probability of dominant-scale breaking, which decreases as the seas age. For old seas, the predominant breaking scale transitions to the shortest whitecaps and non air-entraining microbreakers. The spectral distribution of the breaking dissipation rate contribution to the total dissipation rate in the wave boundary layer is not known with any precision, as direct measurements are presently not feasible.

Understanding the physics and quantifying the total dissipation rate in the upper ocean, including the wave boundary layer, has attracted considerable interest over the past few decades. There has been ongoing debate as to how well this key region is described by turbulent wall-layer scaling, for which the local TKE dissipation rate $\varepsilon_{wl}(z)$ at mean depth z below the ocean surface is given by

$$\varepsilon_{wl}(z) = \frac{u_{*w}^3}{\kappa z}$$

where $\kappa \sim 0.41$ is the Karman constant and $u_{*w}$ is the water-side friction velocity (e.g. Terray et al. 1996)). However, recent consensus strongly favors a breaking-wave enhanced layer where the near-surface TKE dissipation rate exceeds the wall layer estimate by a considerable margin. Section 1 in Sutherland and Melville (2015a) (hereafter SM15a) presents a state-of-the-art account of this extensive literature and also reports the most recent measurements and findings.

Briefly, new insights are emerging as a result of novel measurement techniques and analyses reported in recent field investigations. These studies have reported comprehensive results for a broad range of open ocean wind and sea state conditions that link refined subsurface TKE dissipation rate measurements, novel surface dissipation rate measurements and co-located surface wind and wave properties (e.g. Gemmrich, 2010, Schwendeman et al., 2014, SM15a). This includes spectrally-resolved breaking wave measurements, from which spectral breaking wave dissipation rates can be estimated using Phillips (1985) (hereafter P85) spectral breaking wave framework and its recent refinements (e.g. Romero et al., 2012). Key open questions are being revisited, including the dependence of depth-integrated dissipation rate on wind speed and wave age, with a special focus on the contribution made by breaking waves. Throughout this paper, we use the term wave age to denote the mean wave age ($c_m/u_*$) parameter adopted in SM15a, where $u_*$ is the wind friction velocity and $c_m$ is an integral measure of the wave speed. This wave age was considered by SM15a (see section 2c. in their paper) to be more closely related to the breaking wind-waves than a simple spectral peak wave age, which can be representative of swell. Here, $c_m = g/\omega_m$, where g is gravity and $\omega_m$ is the mean frequency computed from the frequency spectrum $S_{\eta\eta}(\omega)$ as:

$$\omega_m = \frac{\int_0^\infty \omega s_{\eta\eta}(\omega) d\omega}{\int_0^\infty s_{\eta\eta}(\omega) d\omega}$$

In this context, the relative importance of the different scales of breaking waves from large whitecaps to non air-entraining microbreakers has emerged as a new element. Recent observational results on this topic have been reported by Gemmrich (2010), Gemmrich et al, (2013), Sutherland and Melville (2013, 2015a, 2015b) and by Schwendeman et al. (2014). In their recent measurements investigating optimal correlates for active whitecap coverage, Schwendeman et al. (2015) reported lower correlation with TKE dissipation rate than with wind or wave conditions, with residuals showing a strong negative trend with wave age. They suggested that the discrepancy is likely due to the increased influence of microbreaking in older wind seas.

In this paper, we focus on the key issue of the relative contributions of the different breaking wave scales to the total dissipation rate in the wave boundary layer and how this changes for different wave ages. Our study has a special emphasis on the relative importance of the contribution of the very small whitecaps and the non air-entraining microbreakers. This topic is of central importance, as wave breaking is a key air-sea interaction process whose sea surface expression in the form of whitecapping or microbreaking is currently under active investigation. Associated with this phenomenon are several important scientific and potential applications, including fundamental air-sea interfacial fluxes and the utilization of breaking wave signatures as a remote sensing tool for inferring these fluxes. Also, it is essential to know the resolved wave scale bandwidth that needs to be included explicitly or parametrically in models in order to capture the dominant physics in air-sea interaction models.

In this context, by re-analyzing the published data of SM15a, we review their key conclusion regarding the relative importance of the contribution of non air-entraining micro-breakers and very small whitecaps to the total dissipation rate in the wave boundary layer. SM15a propose that the high levels of TKE dissipation rates observed very near the sea surface are consistent with the large fraction of wave energy dissipation rate attributable to non air-entraining microbreakers and very small whitecaps.

## 2. Key results from Sutherland and Melville (2015a)

SM15a combines their novel geometric/kinematic breaking wave crest length spectral density measurements ($\Lambda(c)$) reported in Sutherland and Melville (2013), with parametric spectral breaking strength coefficients from spectral wind wave modeling, to infer the dissipation rate contribution from breaking wave scales from large whitecaps to non air-entraining microbreakers. These results are described and reported in detail in SM15a. While the breaking front imagery allows extraction of directional distributions of $\Lambda(c,\theta)$, the results presented are for the azimuthally-integrated distribution

$$\Lambda(c) = \int_0^{2\pi} c\Lambda(c,\theta)d\theta$$

The integrated dissipation rate contribution from breaking waves of all resolved breaker scales is given by the fifth moment of Λ(c), weighted by the spectral breaking strength coefficient b(c), according to:

$$\int Sds = \frac{\rho_w}{g}\int b(c)c^5\,\Lambda(c)dc$$

(P85, Banner and Morison (2010), Romero et al. (2012), SM15a).

Figures 6, 7 and 16 in SM15a provide the basis for our analysis, as they contain the data directly relevant to the assessment of the relative importance of the microbreaker and short whitecap contributions to the dissipation rate. The data from Figure 7 in SM15a are redrawn in Fig.1 below. This figure shows the cumulative integral of the spectral *breaking* dissipation rate normalized by the total *breaking* dissipation rate for each measured wave age case. It should be noted that this figure only addresses breaking wave dissipation rate contributions, and not the total dissipation rate.

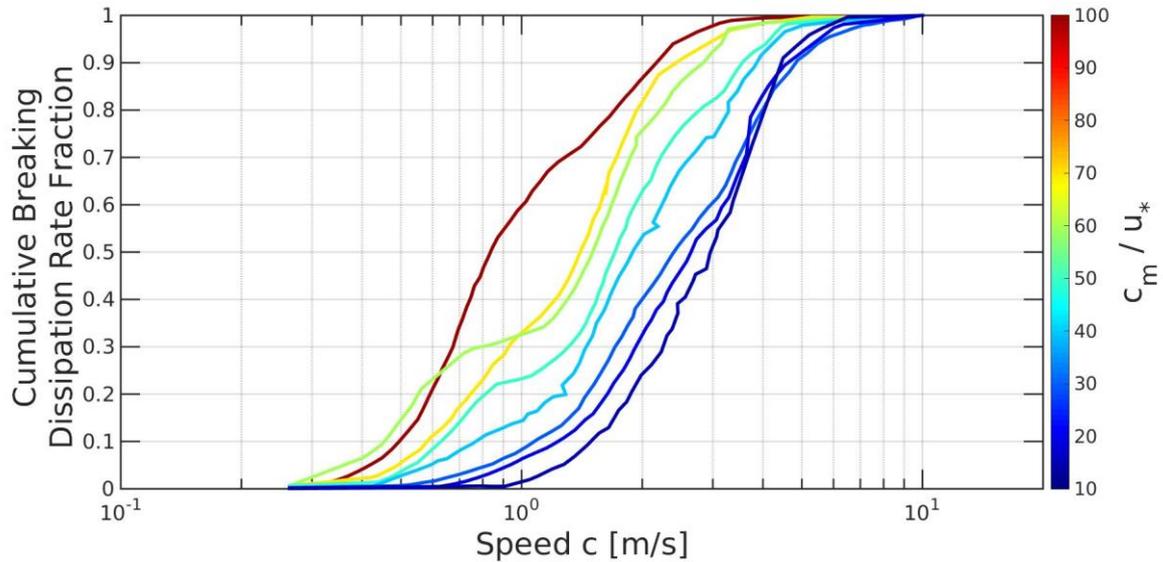

*Fig.1 Cumulative breaking wave dissipation rate normalized by the total breaking wave dissipation rate, plotted as a function of speed of the breaking front (c) for the range of wave age ($c_m/u_*$) conditions indicated in the attached color bar. This distribution reproduces Figure 7 in SM15.*

As seen in Fig.2, there is a 'background' dissipation rate contribution in the wave boundary layer from other hydrodynamical processes. These include the influence of surface waves on the Reynolds shear and normal stresses in the subsurface turbulence, and any resultant energy transfer between the waves and the turbulence. This is important in quantifying the wave energy dissipation rate due to the interaction between non-breaking waves and turbulence, which is a source of wave damping additional to wave breaking. The recent paper by Guo and Shen (2014) includes a comprehensive review of the literature on this topic. In the more complex oceanic context, Sullivan and McWilliams (2010) highlight the need to include Langmuir turbulence, larger wavenumber bandwidth and directional spreading for the surface waves to the non-breaking contributions. In the present paper, the background dissipation rate is taken as that arising from all sources *other than* actively-injected breaking wave turbulence.

The key challenges investigated in this paper are to quantify, as the seas evolve: (i) the fractional contribution by breaking waves to the total dissipation rate; (ii) the relative importance of the microscale breakers and very small whitecaps to the breaking wave and total dissipation rates.

A second key figure underpinning our analysis is Figure16 in SM15a, redrawn as Fig.2 below. This figure shows the measured *total* dissipation rate integrated across the wave boundary layer plotted against the breaking dissipation rate integrated over all resolved wave scales, for a range of wave age conditions from developing wind seas to old swell.

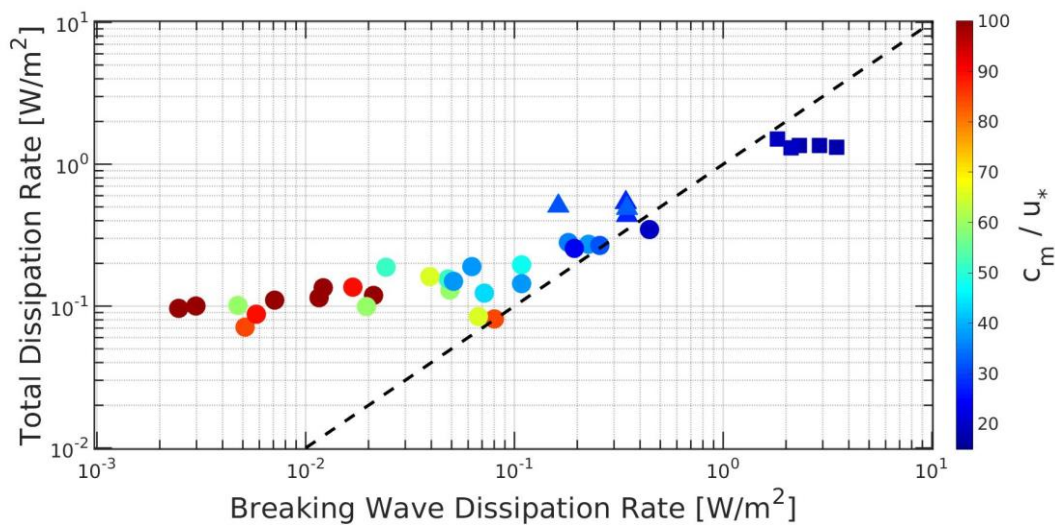

*Fig.2    Measured total dissipation rate integrated across the wave boundary layer plotted against the breaking wave dissipation rate integrated over all resolved wave scales, for the range of wave age ($c_m/u_*$) conditions from developing wind sea to old swell indicated in the attached color bar. This reproduces Figure 16 in SM15a.*

### 3. Re-analysis of the Sutherland and Melville (2015a) results in Section 2.

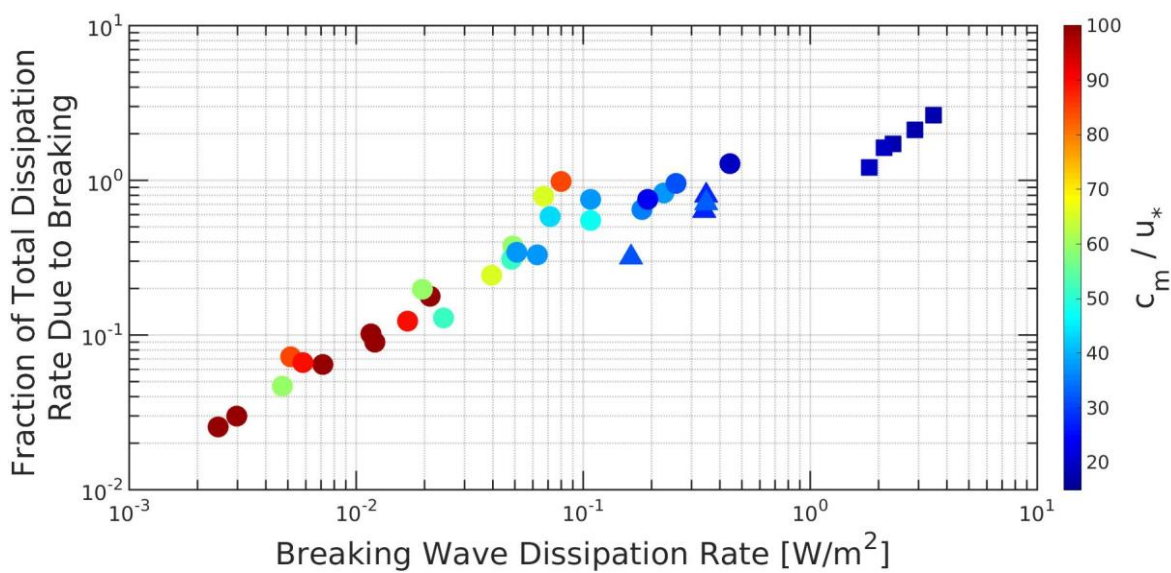

*Fig.3  This figure replots the data in Fig.2 above, retaining the horizontal axis, symbols and colors. The vertical axis now shows the fractional contribution of the breaking wave dissipation rate to the total measured dissipation rate in the wave boundary layer.*

Fig.3 shows that for the high dissipation rates in developing wind seas (right side of plot), wave breaking accounts for almost the entire dissipation rate. However, for the old wind seas (left side of plot), wave breaking contributes only a small fraction of the total dissipation rate. Further, using this data, the fraction of the dissipation rate contributed by wave breaking to the total dissipation rate can be plotted against the wave age, as shown in Fig.4.

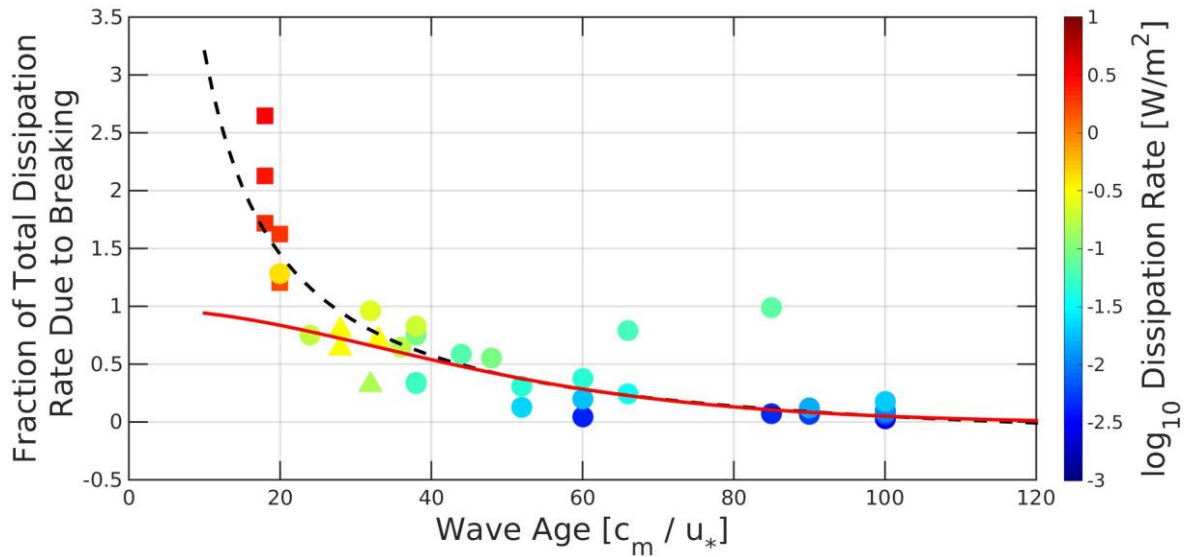

*Fig.4 Fraction of the total dissipation rate from breaking, plotted against wave age $c_m/u_*$. The colors represent $log_{10}$ of the total dissipation rate. This figure is a transformation of the data in Fig.3 above. The solid red and black dashed curves are two fits to the data, as explained in the text.*

In Fig.4 it is seen that for the younger wind seas, the breaking dissipation rate accounts for almost the entire total dissipation rate. However as the seas age, this fraction decreases until it becomes insignificant (< 5 %). The black dashed line in Fig.4 is a linear least squares fit to the data. However, the data (and curve) extend above 1, which is physically implausible. The solid red line is a fit to the dashed black line that asymptotes just below 1.

A key aspect of Fig.1 (Figure 7 in SM15a) is that it does not show the absolute levels for the breaking dissipation rate for each of the wave age cases. However, this information is essential to assess the relative importance of different wave breaking scales to the overall dissipation rate in the wave boundary layer. We were able to extract this information from the b(c)-weighted fifth moment spectra of Λ(c) provided in Figure 6(d) of SM15a, which we digitized and integrated. This figure is redrawn as Fig.5 and has allowed replotting the cumulative breaking wave dissipation rate as a fraction of the total breaking dissipation rate.

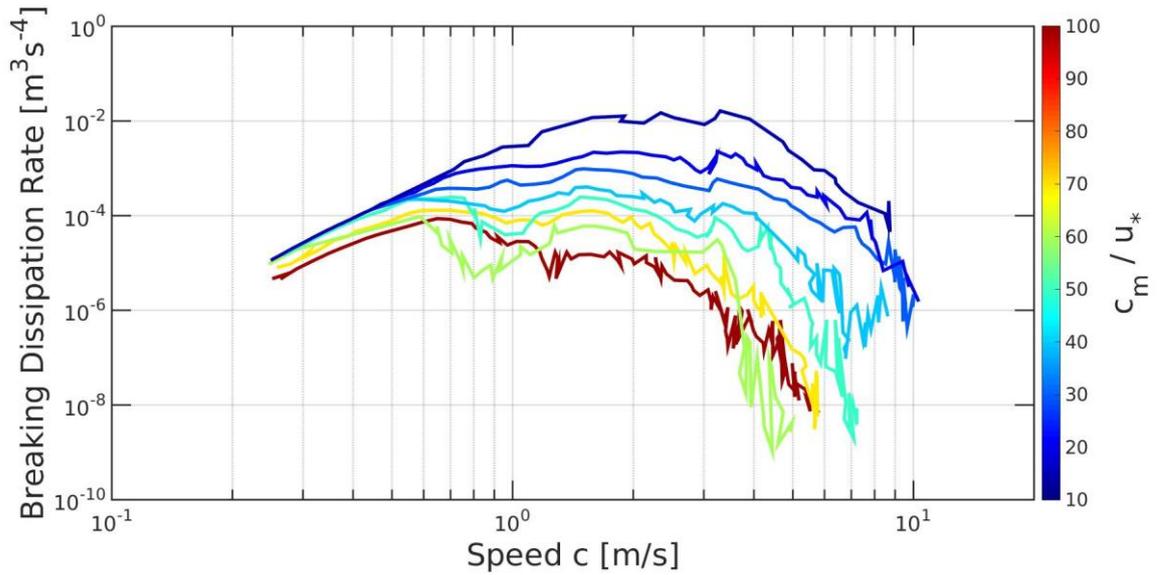

*Fig.5 Spectral breaking dissipation rate against speed c (redrawn from Figure 6(d) in SM15a).*

In SM15a Figure 6(d), the breaking dissipation rate is defined as $S_{dsbr}(c) = b(c) c^5 \Lambda(c)$ [$m^3 s^{-4}$]. In preparing Fig.5, the units were converted to W/m$^2$ by multiplying by $\rho_w/g$, where $\rho_w$ is the water density and g is the gravitational acceleration. In this figure, the breaking crest length spectral density, $\Lambda(c)$, was measured in three different field experiments and the breaking strength coefficient b(c) is a function of c, unlike in P85. Here b(c) is calculated using the modeling strategy proposed by Romero et al. (2012), as described by equation (5) in SM15a. In Fig.5 the line colors and corresponding average wave age bins are the same as used in Fig.1. While this data is quite noisy, it is readily integrated to obtain a reasonably accurate representation of the mean total dissipation rate for each of the average wave age bins in Fig.1.

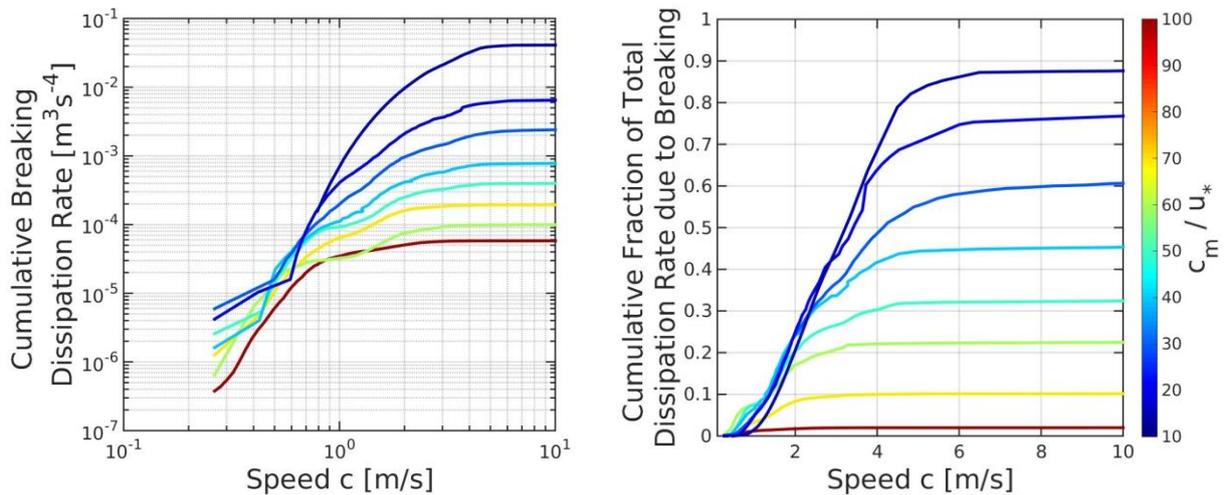

*Fig.6 (a) Cumulative breaking dissipation rate against breaker front speed c for a range of average wave age bins, for the experiments reported in SM15a. (b) Cumulative breaking dissipation rate as a fraction of total dissipation rate plotted against wave speed c. The colors represent average wave age ($c_m/u_*$) for both, and are the same as in Fig.1 and Fig.5.*

Using the data in Fig.6a together with the average $c_m/u_*$ for each bin and the red curve fit in Fig.4, the corresponding average total dissipation rate can be calculated for each of the average wave age bins plotted in Fig.5. The wave breaking dissipation rate can now be plotted as a fraction of the total dissipation rate against breaker front speed c, for the wave age bins in Figs.1 and 5, and is shown in Fig.6b.

Fig.6 demonstrates the following:
(i) for young seas, the total breaking dissipation rate is a large percentage of the total dissipation rate, whereas for very old seas, it is only a very small fraction of the total dissipation rate;
(ii) for young seas, the larger-scale breaking waves make a substantial relative contribution to the total dissipation rate, whereas for very old seas, only the small breaking waves contribute to the total dissipation rate. Hence it can be concluded that for old wind seas, the larger-scale waves have very low breaking probabilities. This can also be seen in Fig.5.
(iii) for old seas, breaking waves only make a small( < 10%) contribution to the total dissipation rate and for young seas, breaking waves traveling at less than 2 m/s make up less than 30% of the total dissipation rate.

To elucidate further the relative contribution to the total dissipation rate made by all breaking wave fronts with speeds below a particular speed, the low wind speed results in Fig.6 are extracted and replotted in Fig.7.

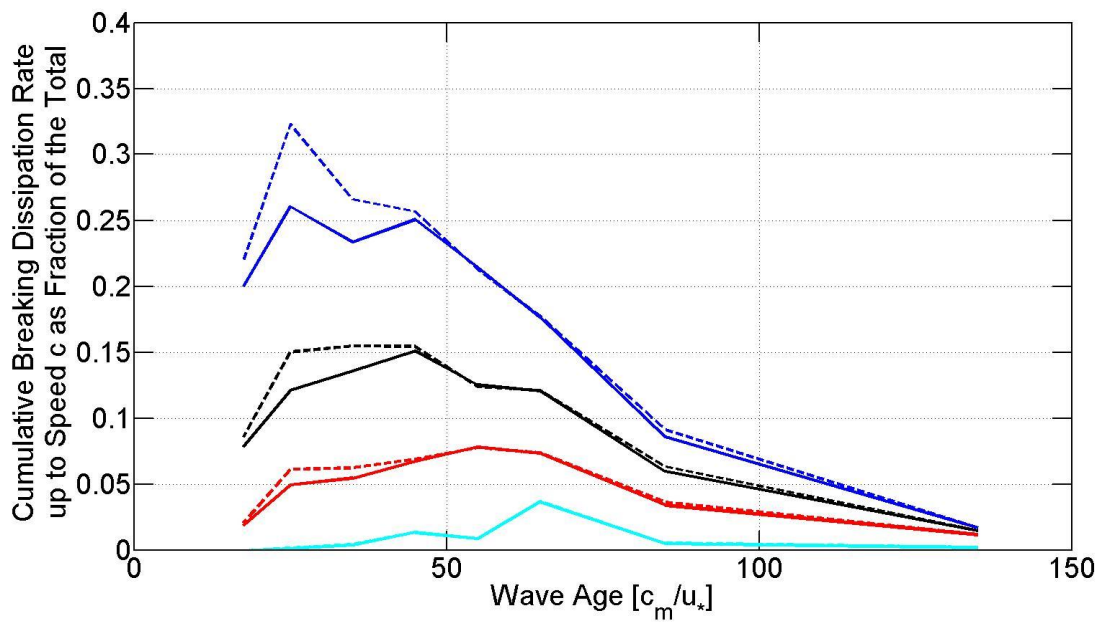

*Fig.7 Cumulative breaking dissipation rate from breaking fronts with speeds up to c m/s as a fraction of the total dissipation rate, plotted against wave age, for c=0.5 (cyan), c=1 (red), c=1.5 (black) and c=2 (blue). The solid and dashed lines are based, respectively, on the solid red line and black dashed line fits to the data in Fig.4.*

From the results in Fig.7, it is seen that microbreaker and small whitecap fronts traveling at speeds less than 0.5, 1, 1.5 and 2 m/s never contribute respectively more than about 3%, 8%, 16% and 26% of the total dissipation rate. This reduces to less than 5% for old seas, where active breaking only plays a small role in the total dissipation rate.

## 4. Impact of alternative Λ(c) extraction methodology

This section reviews the sensitivity of the main findings reported in Section 3 to the methodology used to recover Λ(c) from the whitecap imagery.

### 4.1 *Breaking wave scale*

Three methods for analyzing breaking wave video imagery are currently used to extract Λ(c) from video of breaking waves at sea, with the most appropriate method still to be decided. Each method delivers a mean distribution of breaking crest front length per unit area of sea surface as a function of (vector) wave scale. Measuring breaker length scales directly from video imagery is not straightforward. The preferred method is to measure the breaker crest front velocity, inferring breaker wavelength from the wave speed via the linear deep water wave dispersion relation. This approach introduces other issues, as discussed below.

During active breaking, a breaker crest front slows down to about half its initial velocity (Kleiss and Melville, 2011, Figure 13c; Gemmrich et al. 2013, Figure 1). This intrinsic unsteadiness results in a significant difference in the Λ(c) distributions from each method, because each method assigns a different velocity to a given detected breaking front. We point out that the main purpose of the Λ(c) distribution is to provide the basis for an accurate, unbiased estimate of the wave energy dissipation rate from breaking for each wave scale.

The P85 spectral breaking wave framework uses the breaking front velocity $\mathbf{c}_b$ to characterize the wave scale of each breaker analyzed, but does not address the unsteady aspects of $\mathbf{c}_b$ in building up the Λ($\mathbf{c}_b$) distribution. Since an active breaker develops on a wave crest which travels at speed c, P85 assumed $c_b$ = c, which is then taken as the initial speed at which the whitecap front travels. This is the $c_b$ adopted by Zappa et al., (2012), Gemmrich et al. (2013), amongst others. However, further refinement to relate $c_b$ to c is needed, as discussed in Section 4.2 below.

SM15a used a different approach, described in detail by Kleiss and Melville (2011), which bins the time-dependent history of each breaker front according to its *instantaneous* velocity. This method redistributes the instantaneous Λ(c) contribution from each breaking front during its active lifetime to a bandwidth of slower velocities, down to 50% of its initial velocity. This strategy imparts an irreversible aliasing of the Λ(c) distribution to lower wave speeds, causing a systematic bias of the breaking dissipation rate distribution to shorter wave scales. Assuming the linear gravity wave dispersion relation, this also results in a weighted redistribution of Λ(k) and the spectral breaking dissipation rate contribution for each breaker from its initial breaking wavenumber across shorter scales out to O(4) times its initial wavenumber.

A third method based on whitecap processing in the image spectral (FFT) domain was introduced by Thomson and Jessup (2007) and used subsequently by Thomson et al. (2009). While this method significantly reduces the data analysis effort, it is also influenced by the breaker front slowdown and has windowing issues with aliasing beyond the measured bandwidth (Schwendeman et al., 2014). This method is mentioned for reference only, and hence not investigated here.

### 4.2 *Generic wave crest slowdown*

The generic crest speed slowdown mechanism (Banner et al., 2014) affects the initial breaker front speed and needs to be addressed in the analysis. In his laboratory study of breaking wave packets in a wave basin, Allis (2013) made frame-by-frame video measurements that show the leading edge of a spilling breaker advances unsteadily at a mean rate less than the equivalent linear wave speed $c_0$ of the underlying wave. The average speed of all 2D and 3D breakers generated by the snake paddle in his wave basin was found to be $[0.87 \pm 0.08]c_0$. This observed generic breaking front speed of ~$0.87c_0$ needs to be taken into account when assigning the correct speed scale c for the $\Lambda(c)$ extracted from video measurements. The observed 13% mean speed reduction requires a 15% increase (1/0.87=1.15) in the speed attributed to a given breaking event to match the linear speed of the underlying wave. This will displace the observed $\Lambda(c)$ distribution towards higher speeds. Note that this correction is also crucial if the data is subsequently transformed to $\Lambda(k)$ distributions for comparison with standard spectral wave model output. This results in a marked systematic shift to O(30%) lower wavenumbers.

After applying the generic slowdown correction determined by Allis (2013), we now suppress the b subscript in $c_b$ and adopt c as the appropriate speed-corrected measure of the wave scale, with the corresponding spatial wavenumber k given by the linear dispersion relation.

4.3  *Impact on $\Lambda(c)$ spectra and resultant breaking dissipation rates*

This subsection quantifies the impact of the two factors in 4.1 and 4.2 above on the $\Lambda(c)$ and allied results shown in section 3 of SM15a. It will be seen that the processing methodology has a potentially strong influence on the results and subsequent conclusions regarding breaking dissipation rate contributions.

The starting point for our analysis is the set of $\Lambda(c)$ distributions reported in Sutherland and Melville (2013) derived from their video and infrared video image data recorded from RP FLIP and processed by their analysis methodology. This data set shows representative $\Lambda(c)$ spectra obtained for a range of banded mean wave age conditions $c_m/u_*$ ranging from 20-170. These spectra are shown in Figure 6(a) of SM15a. As highlighted above, the hallmark of their processing is that during each analyzed breaking event in the ensemble captured by the video, the instantaneous breaker crest front length is binned according to its instantaneous front speed. The ensemble of these individual partitionings makes up the distribution from which each of their $\Lambda(c)$ spectra are created. Full details are given in Sutherland and Melville (2013) and its supplementary annexe.

Our first goal was to reconstruct, as closely as possible, the $\Lambda(c)$ spectra consistent with the P85 framework, hereafter labeled $\Lambda_{BL}(c)$, denoting Base Line $\Lambda$, prior to the subsequent instantaneous speed bin (hereafter ISB) transformation applied by SM15a for each breaking event. The ISB-transformed $\Lambda(c)$ reported by SM15a, hereafter labeled $\Lambda_{IS}(c)$, cannot be inverted explicitly, so we applied an iterative ISB transformation to a set of trial $\Lambda_{BL}$ spectra to emulate the SM15a methodology described above. This Transformed Baseline $\Lambda$ is labeled $\Lambda_{TB}(c)$. If the initial guess for the $\Lambda_{BL}(c)$ is correct, then the $\Lambda_{TB}$ will closely match the $\Lambda_{IS}$. The trial spectral function $\Lambda_{BL}(c)$ before modification by the SM15a ISB process was modeled as having the form:

$$\Lambda_{BL}(c) = r(c)\ c^{-n} \tag{5}$$

where r is an assumed spectral roll-off function at slow wave speeds and $c^{-n}$ is an assumed power law fall-off that follows the measured overall trend. Forms for r and values for n were refined by successive iteration so that the ISB-transformed $\Lambda_{TB}(c)$ spectra matched as closely as possible the $\Lambda_{IS}(c)$ spectra reported by SM15a derived from their measurements.

For each wave age banded case, we formulated a base $\Lambda_{BL}(\mathbf{c})$ spectral function (corresponding to the P85 framework, as specified above (equation (A1)). Based on the measured space-time properties of spilling breakers reported in Figure 13 of Kleiss and Melville (2011), the matching procedure consisted of applying a cosine-weighted spectral function (filter) which spreads the initial speed over a speed range from the initial speed to half the initial speed, peaking at about 0.75 of the initial speed. Our equivalent spectral weighting starts at the fastest moving scale $c_{max}$ and redistributes the spectral density by a windowing function that varies cosinusoidally, varying proportionally from 0.5 at $c_{max}$, to 1 at $0.75c_{max}$ and back to 0.5 at $0.5c_{max}$. These weightings were then normalized so that their sum was 1. Applying this window successively to our base spectrum $\Lambda_{BL}(c)$ starting at $c_{max}$ and moving sequentially towards slower speeds reshapes to lower speeds the initial base spectrum $\Lambda_{BL}(c)$ to the transformed baseline spectra $\Lambda_{TB}$ for comparison with the $\Lambda_{IS}(c)$ spectra shown Figure 6(a) of SM15a. By successive iteration in which the exponent n and the parameters in the function r (eqn. (1)) are tweaked, a close correspondence was achieved between the $\Lambda_{TB}(c)$ and the corresponding $\Lambda_{IS}(c)$ spectrum reported for each different wave age band in SM15a. We also tried other weighting functions, such as uniform weighting, which spread the speed over 0.5 to 1. However this only marginally altered the result.

Further, from Allis (2013, Ch.7, p.190, below Figure 7.10), the observed breaker crest front speed c=|**c**| is only 0.87 of the actual underlying linear wave speed, and so the c dependence in $\Lambda_{BL}(c)$ needs to be replaced by $c=1.15c_0$ to match it to the actual wave speed of the underlying wave, as described towards the end of the Overview section above. This speed correction is also applied to the $\Lambda_{BL}$ spectra to produce a speed-corrected baseline spectrum which is referred to as $\Lambda_{BC}$ (for Baseline Corrected). Fig.8 shows a typical example of the four different spectra $\Lambda_{IS}(c)$, $\Lambda_{BL}(c)$, $\Lambda_{TB}(c)$ and $\Lambda_{BC}(c)$.

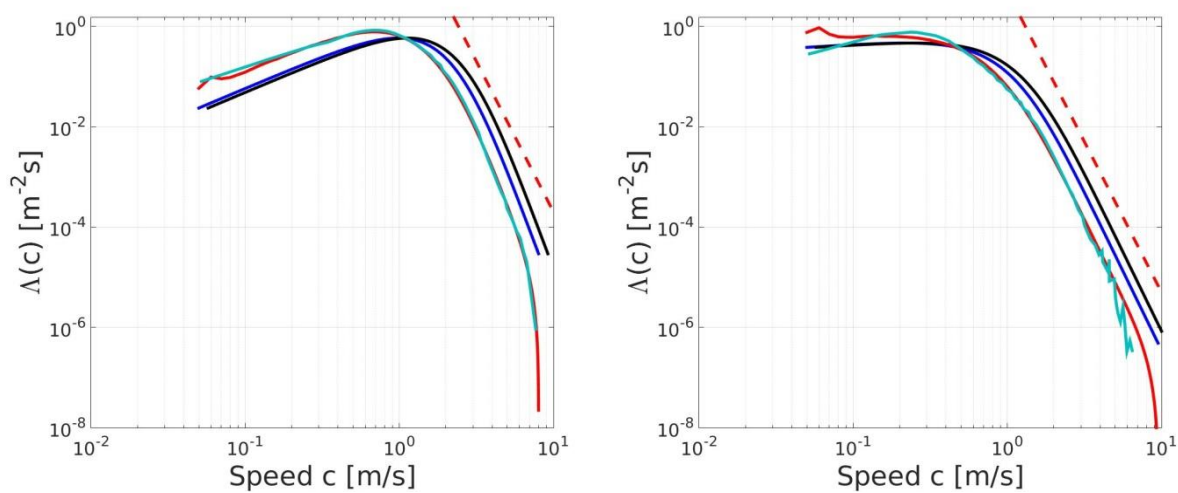

*Fig.8. Left Panel is for young seas, right panel is for old seas. Both panels are plots of $\Lambda(c)$ against c. The cyan line is $\Lambda_{IS}(c)$, the blue line is $\Lambda_{BL}(c)$, the red line is $\Lambda_{TB}(c)$ and the black line is $\Lambda_{BC}(c)$. The red dashed line is the P85 $c^{-6}$ dependence.*

From Fig.8 it is seen that the fit between $\Lambda_{IS}(c)$ and $\Lambda_{TB}(c)$ is good for young seas and for most speeds for the older seas. The noise in the observed $\Lambda_{IS}(c)$ for larger values of c in older seas arises from a combination of low breaking probabilities in the peak region for older seas, and a limited-duration data record in the observations. Also, $\Lambda_{IS}(c)$ shows a significant shift of the spectral peak towards shorter, slower waves relative to the baseline spectrum $\Lambda_{BL}(c)$, and its spectral peak level has increased. This results from the speed-binning of the slowing breaker fronts. For each of these spectra, the spectral fall-off towards faster waves follows the reference $c^{-6}$ dependence predicted by P85. Also note that all the $\Lambda(c)$ the peak sea values for the old seas are much lower than for the young seas.

As the fifth moment $c^5\Lambda(c)$ underpins the breaking dissipation rate (SM15a, equation(4)), the recovered four different $\Lambda_{nn}(c)$ are multiplied by $c^5$, to investigate the affect of the above transformations on the breaking dissipation rate. The matching plots are shown in Fig.9.

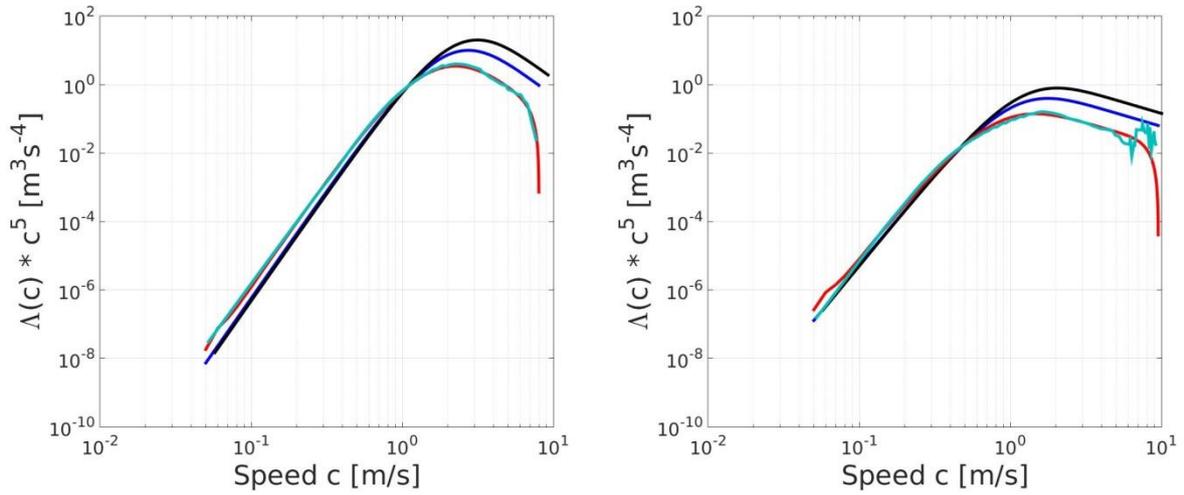

*Fig.9. Same as Fig.8, but for the fifth moments of the spectra $\Lambda_{IS}(c)$, $\Lambda_{BL}(c)$, $\Lambda_{TB}(c)$ and $\Lambda_{BC}(c)$.*

The $c^5$ weighting has changed the spectral peak level differential between $\Lambda_{IS}(c)$ and $\Lambda_{BL}(c)$, with the latter now exceeding the former. Note that after speed-correction of $\Lambda_{BL}(c)$ to $\Lambda_{BC}(c)$, the difference between $\Lambda_{BC}(c)$ and $\Lambda_{IS}(c)$ is about one order of magnitude for this case.

Finally, to obtain the breaking dissipation rate, the data in Fig.9 was multiplied by the breaking strength parameter b(c) used in SM15a, formulated following Romero et al. (2012). While we do not have the direct data for b(c), we have $\Lambda(c)$ versus c for all the wave age bins from SM15a Figure 6a, and $b(c)*c^5*\Lambda(c)$ for all the wave age bins from SM15a Figure 6d. By combining the information from these 2 figures we can extract b(c) for all the wave age bins.

We then plot the breaking dissipation rates based on the different $\Lambda$ spectra using the P85 relationship, shown in Fig.10.

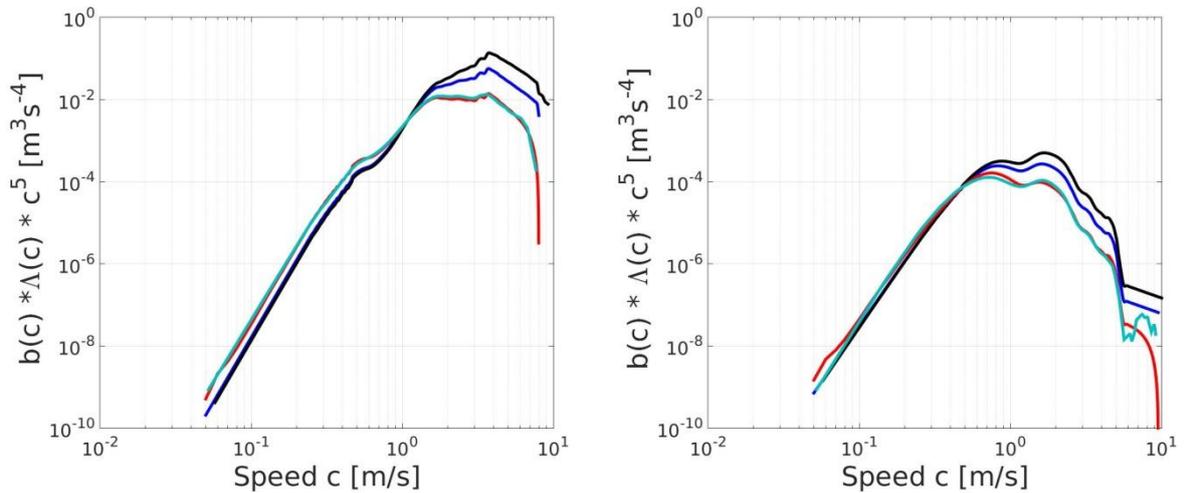

*Fig.10. The spectral breaking dissipation rate for each of the $\Lambda$ distributions in Fig.8, with young seas on the left and old seas on the right. Colors are the same as in Fig.8 & Fig.9.*

From Fig.10 it is seen that for the young seas, the peak breaking waves dominate the breaking dissipation rate for all four $\Lambda$ distributions, with the $\Lambda_{BC}(c)$ breaking dissipation rate higher in the peak region than $\Lambda_{IS}(c)$, but lower for the smaller scale breakers. This is also true for the old sea case, except spectral peak breaking waves no longer dominate the breaking dissipation rate. The older sea peak wave breaking dissipation rate values are much lower than for the young seas.

The cumulative breaking dissipation rate results from $\Lambda_{IS}(c)$ in Fig.1 and $\Lambda_{BC}(c)$ in Fig.9 can also be compared by constructing the corresponding normalized cumulative versions of the results shown in Fig.10. These are plotted in Fig.11.

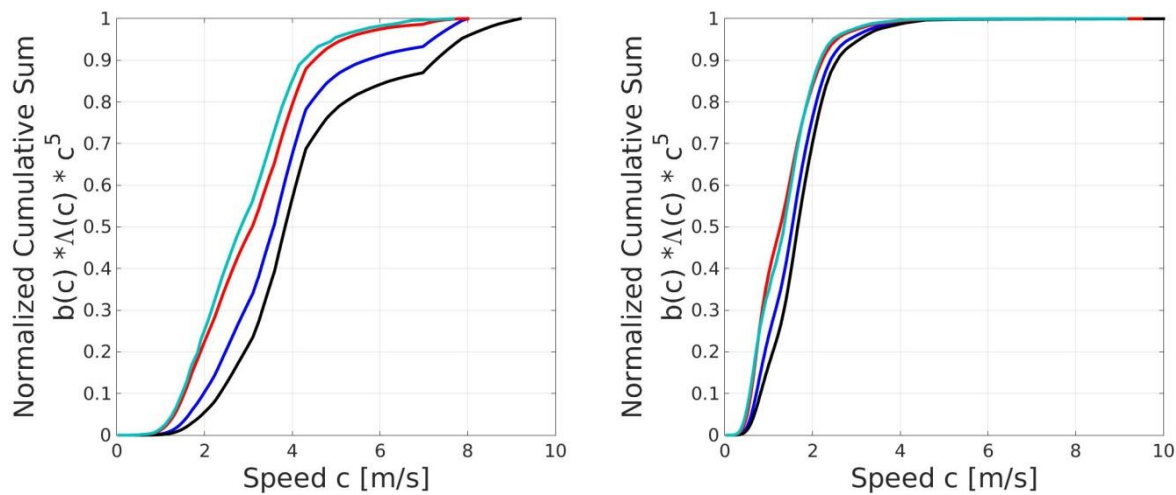

*Fig.11. The normalized cumulative results of the data shown in Fig.10. Young seas are shown on the left and old seas on the right. Colors are the same as in Figs.8, 9 & 10.*

In Fig.11, it can be seen that for young seas (left panel) the cumulative dissipation rate from breaking waves traveling at less than 2 m/s is a small percentage of the total breaking wave dissipation rate, and that the different $\Lambda$ processing methods have a notable affect. For the older sea case in the right panel, the breaking waves traveling at speeds less than 2 m/s contribute a large percentage of the total breaking wave dissipation rate. It is worth re-emphasizing that this is only the dissipation rate from wave breaking, and that for old seas the dissipation rate from breaking is only a small fraction of the total dissipation rate.

4.4 *Implications for the spectral dissipation rate contributions from breaking waves*

The impact of the above considerations regarding $\Lambda(c)$ on the allied breaker dissipation rate results reported in SM15a is now investigated in detail. We will show that the contribution of the very short breaking waves to the total dissipation rate is diminished very appreciably relative to that claimed by SM15a if an alternative processing methodology for $\Lambda(c)$ is used that is based on the P85 prescription described in Section 4.1 above.

We determine breaking dissipation rates from the $\Lambda(c)$ spectrum and the spectral breaking strength coefficient b(c), using the methodology given by equation (4) in SM15a. Here, the same b(c) as in SM15a is used. We compare results using the fifth moments of $\Lambda_{IS}(c)$, $\Lambda_{BL}(c)$ and $\Lambda_{BC}(c)$.

Fig.12a shows the normalized cumulative dissipation rate due only to breaking as a function of mean wave age based on our $\Lambda_{BC}(c)$ spectrum, for comparison with the SM15a results based on their $\Lambda_{IS}(c)$ spectrum shown in Fig.1 above.

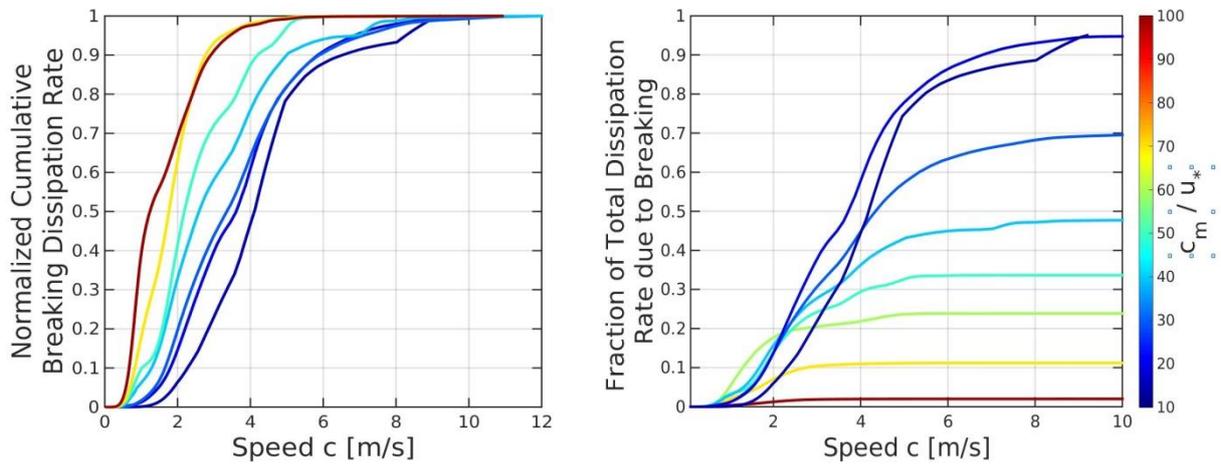

*Fig.12 (a) Normalized cumulative breaking dissipation rate based on the speed-corrected baseline spectrum $\Lambda_{BC}(c)$ against breaker front speed for the different binned wave ages, as in Fig.1. (b) Cumulative breaking dissipation rate as a fraction of the total dissipation rate, calculated from the speed-corrected baseline spectrum $\Lambda_{BC}(c)$ for breaking fronts with speeds up to speed c, for the range of binned wave age cases reported in SM15a.*

The difference between Fig.12a and Fig.12b is that Fig.12a is normalized by the spectrally integrated dissipation rate due to breaking, whereas Fig.12b is normalized by the total dissipation rate from all sources. For old seas (red line) it is seen in Fig 12a that the breaking is dominated by waves traveling slower than 2 m/s, but Fig.12b shows that the breaking dissipation rate is a very small fraction of the total. For younger seas (blue lines) Fig.12b shows that breaking is the dominant form of dissipation, but waves traveling slower than 2 m/s make up, at most, less than 20% of the total dissipation rate. This result is more clearly shown in Fig.13.

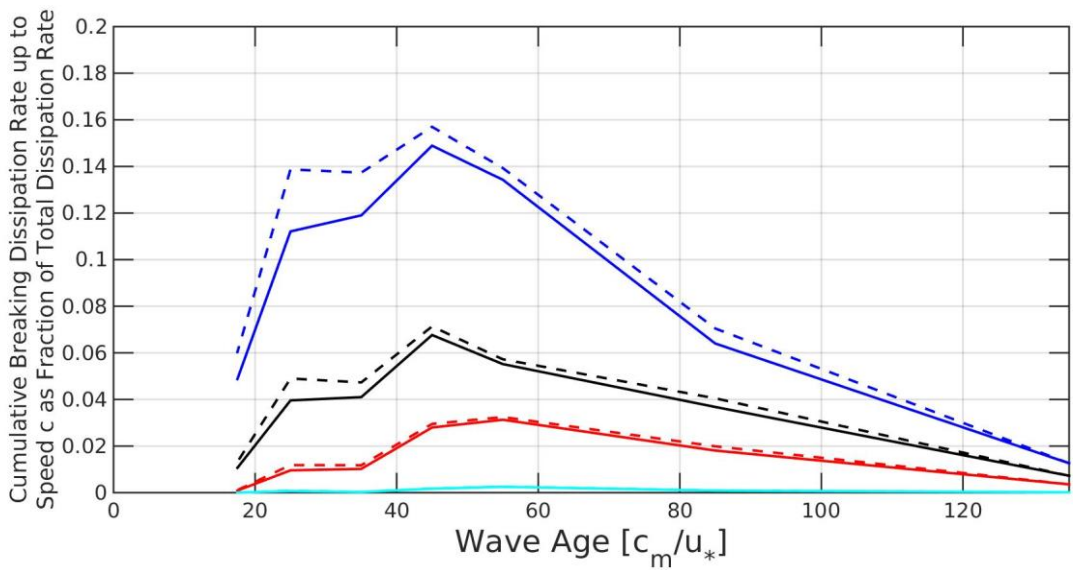

*Fig.13 Cumulative breaking dissipation rate based on $\Lambda_{BC}(c)$ from breaking fronts with speeds up to c m/s as a fraction of the total dissipation rate, plotted against wave age, for c=0.5 (cyan), 1 (red), 1.5 (black) and c=2 (blue). The solid and dashed lines are based respectively on the solid black and red line fits to the data in Fig.4.*

When Fig.13 based on $\Lambda_{BC}(c)$ is compared with Fig.7 based on $\Lambda_{IS}(c)$, it is seen that the magnitude of the contributions in Fig.13 is almost halved relative to Fig.7. This indicates an even lower contribution of the microbreakers and small whitecaps to the total dissipation rate in the wave boundary layer.

4.5 *Sources of uncertainty*

In the complex suite of measurements and analyses utilized by SM15a in their detailed investigation, there are a number of sources of uncertainty that underpin the results. These are discussed in SM15a or in various allied papers by these authors or their collaborators. For example, Fig.16 in SM15a documents the uncertainty in their subsurface TKE dissipation rate measurements. Fig.10 in Sutherland and Melville (2015b) provides uncertainty estimates associated with the estimation of the surface TKE dissipation rate. Kleiss and Melville (2011) discuss various uncertainties associated with extracting crest length spectral density distributions from sea surface visible imagery, with Sutherland and Melville (2013) providing its counterpart for infrared imagery. This aspect includes the contentious issue of breaking front speed assignment, for which the uncertainty is investigated in detail in the present paper. In estimating spectral breaker energy dissipation rates, a spectral breaker strength formulation is needed. This quantity has a significant uncertainty, as is evident from the paper by Romero et al. (2012). We avoided this issue in the present study by using the same spectral breaking strength data used by SM15a, which was based on Romero et al. (2012). However, this quantity remains a source of significant uncertainty with a potentially large impact on breaking dissipation rate estimates, as pointed out in the discussion of Fig.4 above. In any event, our study has been carried out so that our conclusions below are not influenced by these various uncertainties.

## 5. Conclusions

We revisited the findings of the recent SM15a (Sutherland and Melville (2015a)) field study, especially the relative importance of the contributions of microbreakers and very small-scale whitecaps to the total dissipation rate in the oceanic wave boundary layer.

SM15a found that measured integrated total dissipation rates in the water column agreed well with measured dissipation rates from breaking for wave age conditions ($c_m/u_* < 50$), beyond which breaking contributions become secondary to the background dissipation rate.

However, recent infrared $\Lambda(c)$ measurements of Sutherland and Melville (2013) showed that previous field measurements of breaking waves failed to capture non air-entraining microbreakers. These novel $\Lambda(c)$ measurements were used by SM15a to estimate breaking wave energy dissipation rates and to conclude that a large fraction of wave energy was dissipated by microbreakers and small whitecaps.

We reviewed the data analysis methodology used by SM15a to arrive at their conclusions. Based on our analysis methodology, we find that the contribution of these very short breakers to the total dissipation rate during active wind-wave generation conditions is far weaker than claimed by SM15a, and is only ever of marginal importance.

More specifically,

- for young/developing wind seas, the very short breaker contribution to the total dissipation rate is small compared with the contribution from the larger-scale breakers
- for low wind speeds/very old seas, the small-scale breakers dominate the breaking wave dissipation rate contribution. However, measured background total dissipation rate measurements are an order of magnitude larger than the total breaking-induced contributions.

Hence, to determine total dissipation rates in the wave boundary layer, the breaking dissipation rate from microbreakers and very small whitecaps is not needed. These contributions are eclipsed by differences in breaker image processing techniques and uncertainty in the spectral breaking strength coefficient, which combined can contribute O(2) uncertainty to the breaking dissipation rates, as discussed in some detail in this paper.